# ENERGY AND PERFORMANCE EVALUATION OF REACTIVE, PROACTIVE, AND HYBRID ROUTING PROTOCOLS IN WIRELESS MESH NETWORK

Jean Louis Ebongue Kedieng Fendji[1] and Sidoine Djuissi Samo[2]

[1]Department of Computer Engineering, University Institute of Technology, Ngaoundéré
[2]Apple College, Dubai, United Arab Emirates

## ABSTRACT

*This paper evaluates the energy consumption of well-known routing protocols, along with other metrics such as throughput, packet delivery ratio (PDR), and delay in different scenarios. We consider two other metrics in order to capture the efficiency of the energy consumption: e-throughput which is the ratio between the consumed energy and the throughput; and the e-PDR which is the ratio between the consumed energy and the PDR. We compare four routing protocols: AODV, OLSR, and HWMP in Reactive and Proactive modes. The number of nodes is varying between 25 and 81 nodes, with different mobility models. Simulations are conducted using NS3 and the parameters of a real network interface card. From the results, AODV presents the lowest energy consumption and a better e-Throughput. OLSR provides a better e-PDR in mobile scenarios. With a smaller e-PDR and e-Throughput, the proactive mode of HWMP is more energy efficient than the reactive mode.*

## 1. INTRODUCTION

Since their introduction more than two decades ago, wireless networks are presented as an appealing solution to connect devices especially in difficult-to-wire areas. Based on an infrastructure or not, they are more preferred than wire networks. When the wireless network is not based on a central infrastructure, it is said to be in ad hoc mode. A wireless ad hoc network is self-organized, that means it can reconfigure itself when a node joins or leaves the network. When nodes are mobile, we talk about mobile ad-hoc networks usually shortened MANETs [1]. Communication between a pair of nodes in a MANET can be done by sending messages through a set of intermediate nodes, which may act as routers. When a node fails, communication in a MANET can still continue as long as the rest of the network is not partitioned. This characteristic provides some robustness to the network. MANETs can be considered as a type of wireless mesh networks (WMN), more precisely client mesh network [2].

During the communication, the path followed by a message in a MANET or a WMN depends on the routing protocol mainly defined at the network layer. Depending on their routing policies, routing protocols can be classified into three types: reactive, proactive, and hybrid. The latter type encompasses two modes: reactive and proactive modes. Routing policies play a central role on the energy consumption and the performance of the network, which can be also influenced by the topology (dynamic or not) of the network.

Quite a lot of routing protocols as well as performance factors have been proposed by the IETF's MANET working group. However, related energy factors have not be considered [1]. The energy consumption is therefore still a crucial issue, regardless the lot of works which has been focused on the improvement of routing protocols performance. This issue is more critical when nodes are powered essentially by a limited source, as it is the case in most scenarios in MANETs and in some recent scenarios in WMNs. In fact, mesh routers whose had fixed locations and were powered by grid energy, have started to be equipped with batteries like in robots [3]. In addition, most WMNs in rural or remote regions are powered by generator or solar panels [4][5][6][7]. These trends rise the crucial need of studying not only the performance factors of the routing protocols in MANETs and WMNs, but also their energy consumption. Despite the plethora of works focusing on the evaluation of routing protocols performance in wireless networks, only few works have been devoted to the estimation of energy consumption [8][9]. An interesting survey on energy-efficient routing schemes for MANETs is provided in [10].

This paper provides an evaluation of the energy consumption and performance metrics of three types of routing protocols: AODV (reactive), OLSR (proactive), HWMP (hybrid) in reactive and proactive modes. In addition, to portrait the energy-efficiency of selected routing protocols, two other metrics are used: e-throughput determined by the ratio between the consumed energy and throughput, and the e-PDR determined by the ratio between the consumed energy and the PDR.

The rest of the paper is organized as follow: the selected routing protocols are presented in Section 2. Section 3 presents earlier works on the evaluation of performance and energy consumption in routing protocols. The simulation parameters and scenarios are described in section 4. Simulation results are presented and discussed in section 5.

## 2. SELECTED ROUTING PROTOCOLS

Three type of routing protocols have been selected for the comparison: AODV as reactive protocol, OLSR as proactive protocol, and HWMP in reactive and proactive mode as hybrid protocol.

### 2.1. AODV

AODV (Ad hoc On-Demand Distance Vector) remains the most popular routing protocols among all the reactive ones. It is inspired from distance vector protocol such as DSDV and DSR and serves as a basis for several other reactive routing protocols. It has been defined in RFC 3561[12]. It enables dynamic, self-starting, multihop routing between mobile nodes in an ad hoc network.

AODV starts with a route discovery process before transmitting data. The route discovery process determines a unicast route to the destination. During this stage, a route request RREQ packet is flooded from the sending node to its neighbours. Each of its neighbours which receive this packet forwards this packet to their neighbours until the destination is found. Once the destination is reached, a route reply RREP packet is sent by the initial sender by considering the route to the source contained in the RREQ packet. This packet follows the reverse path taken by the RREQ, and the route to the destination is updated in all intermediate nodes. The discovery process ends when RREP reaches the initial sender. Data transmission can therefore start. The operation of AODV is loop-free due to the use of destination sequence number as described in [12].

A link may be broken, resulting in an error during data transmission. In this case, the affected set of nodes is notified so that they are able to invalidate the routes using the lost link. To achieve this, a route error RERR packet is sent. Moreover, routes that are not in active communication are not maintained: it is the maintenance process.

## 2.2. OLSR

OLSR (Optimized Link State Routing protocol) is the most popular proactive routing protocol. It has been defined in its first version in RFC 3626 [13] in 2003 and later in RFC 7181 in 2014. In its first version, which is also the most implemented, the route is build beforehand for data transmission by maintaining a routing table at each node. OLSR make therefore use of the following mechanisms as described in [13]:

- Link Sensing: it checks the connectivity between nodes by sending periodic HELLO messages over the interfaces through which connectivity is checked.
- Neighbour detection: In a network with only single interface nodes as it is our case, the neighbour set of a node may be deducted from the information exchanged as part of link sensing.
- MPR Selection and MPR Signalling: Multipoint relays (MPRs) nodes are a set of special nodes selected by each node in its neighbourhood. When a node generates a broadcast message, it is retransmit only by MPRs, in such a way that this message will be received by all nodes 2 hops away.
- Topology Control Message Diffusion: OLSR being a table-driven routing protocol, the routing table at each node is constructed using topology control by the means of Topology Control (TC) packets. Those TC packets are forwarded only by MPR.
- Route Calculation: The routing table at each node, containing sufficient link-state information, will be used for route calculation. The link state information is acquired through periodic message exchange, or through the interface configuration.

## 2.3. HWMP (R and P)

HWMP (Hybrid Wireless Mesh Protocol) has been defined in IEEE 802.11s and dedicated to Wireless Mesh Networks [14]. It supports two modes of operation depending on the configuration: reactive mode and proactive mode. It makes use of four types of control messages: Route Request (RREQ), Route Reply (RREP), Root Announcement (RANN), and Route Error (RERR).

HWMP is essentially a reactive protocol to which a proactive mechanism has been added in order to permit a node to announce itself as the root of a tree-based topology. In this paper, we consider both operation modes. When a source mesh point (MP) needs to find a route in reactive mode, it broadcasts a RREQ indicating a destination MP and the metric field being initialized to 0. A MP creates a route to the source or updates its current route when it receives a RREQ. When a new route is created or an existing route is modified, the RREQ is forwarded. Each MP may receive multiple copies of the same RREQ coming from the source, but each copy has a unique path from the source to the MP. The destination MP sends a unicast RREP back to the source after creating or updating a route to the source. Two important mechanisms are defined: "Destination Only" (DO) flag which indicates whether intermediate MP can generate a RREP; and "Reply and Forward (RF)" flag which indicates whether an intermediate MP can forward a RREQ. When intermediate MPs receive the RREP, they create a route to the destination, and then forward the RREP toward the source. The source creates a route to the destination on receiving the RREP.

## 3. RELATED WORK

Studies on performance and energy consumption evaluation in wireless networks can be classified in two groups: protocols improvement and protocols comparison.

### 3.1. PROTOCOLS IMPROVEMENT

Gupta et al. in [15] proposed Energy-Aware AODV (EA-AODV) routing protocol and compared this extended version to pure AODV in the perspective of remaining energy. The new routing protocol has been found more energy-efficient than pure AODV. Energy Dependent DSR (EDDSR) has been proposed in [16]. This DSR-based routing protocol has been compared with MDR, LEAR and pure DSR. Simulations have been conducted in NS2, using sparse and dense network scenarios. From the results, EDDSR and MDR outperformed pure DSR routing protocol on the basis of node lifetime, and especially in dynamic scenarios. A non-neglecting observation was the higher energy expenditure of LEAR due to its route discovery process especially in dense networks.

Kim and Jang proposed New-AODV, an Enhanced AODV Routing Protocol, which attempts to extend the entire network lifetime by adjusting RREQ delay time [17]. Simulation on NS2 showed the superiority of New-AODV over pure AODV routing protocol.

In [18], authors proposed a new mechanism of Local energy-aware named LEA-AODV for Ad-hoc which is based on the reduction of the energy consumption during the route discovery and the route maintenance phases. From the results, in most of the simulated scenarios LEA-AODV performs better than pure AODV.

Sahnoun et al. in [19] proposed an extended version of OLSR named Energy Efficient and Path Reliability OLSR (EEPR-OLSR). Compared to standard OLSR, EEPR-OLSR provided a better network lifetime and PDR.

Authors in [20] proposed an enhanced version of AODV named AODVCS. It is inspired by the cuckoo search method and implemented in NS2. AODVCS provides similar PDR as AODV; but with substantially low end-to-end delay.

### 3.2. PROTOCOLS COMPARISON

Simulation models for the evaluation of wireless networks performance have been introduced by J. Broch et al. [21] members of the CMU monarch group. They considered three metrics (packet loss, routing overhead and route length) and focused their work on four routing protocols: The Dynamic Source Routing (DSR) [22], AODV, The Temporally-Ordered Routing Algorithm (TORA) [23], and the Dynamic Destination-Sequenced Distance-Vector Routing (DSDV) [24]. Authors in [25] also compared the same routing protocols with a regard on energy consumption. Their work was carried out using Network Simulator-2 (NS2). They observed that AODV and DSR perform better than DSDV and TORA. The poor performance of TORA has been justified in [26] by its inefficient implementation in NS2. AODV, DSR, DSDV, and TORA have been also studied in [27] with a focus on the mobility impact on energy consumption. This study revealed that reactive protocols are more speed-sensitive than proactive ones, apart from scenarios where nodes move in groups. In those later cases, on-demand protocols perform better than proactive ones on the perspective of energy conservation. One of the first works comparing Optimized Link State Routing Protocol (OLSR) and DSR with regard on energy consumption dates back to Fotino and al. [28]. Their main observation was twofold: in low traffic rate, DSR takes advantage of its routing policy, and in higher rate, OLSR consumes less energy. Several works tried to reduce the energy consumption of OLSR later [29], [30].

Authors in [31] compared three routing protocols: AODV, DSR, and OLSR. They defined several scenarios with dynamic topologies based on Random Waypoint mobility model and different number of nodes, up to 15. The main observation is that AODV is less energy-efficient than OLSR, which in turn consumes more energy than DSR, especially in transmission and receiving mode.

Modified versions of AODV have been compared using energy related metrics by Cao in NS2 [32]. He derived six other protocols: Minimum Total Transmission Power Routing (MTPR), Minimum Battery Cost Routing (MBCR), Min-Max Battery Cost Routing (MMBCR), Time Delay On-demand Routing (TDOD), Minimum Drain Rate (MDR), and Conditional Max-Min Battery Capacity Routing (CMMBCR). The main funding is that MTPR is better than other protocols by finding the minimum energy cost path.

GlomoSim Simulator has been used in [33] to evaluate the performance based on energy consumption of AODV, LAR1, and DSR protocols in high density Ad Hoc networks. One of the findings is that LAR1 performs better than the others for high density networks (around thousands of nodes).

In [34], authors presented the energy consumption of DSR and AODV under Self-Similar traffic (Pareto and Exponential) in comparison with Constant Bit Rate (CBR). Simulation conducted using NS2 showed that AODV is better than DSR only in large area shape using little number of nodes. A similar work has been conducted by Kafhali et al. in [35]. Authors compared AODV, DSR and DSDV on the point of view of the total consumed energy and under three mobility models (Random Waypoint Model, Reference Point Group Model, and Manhattan Grid Model). They also considered three traffic models, namely CBR, Exponential, and Pareto. The key result is that AODV is less energy-efficient with CBR traffic when comparing to DSR and DSDV. However, AODV is more energy-efficient when using Pareto and Exponential traffics.

Maan and Mazhar considered different mobility models while evaluating AODV, DSR, DSDV, OLSR, and DYMO (Dynamic MANET on demand) [36]. They took into account well known metrics such as delay, PDR, and normalized routing load, without considering the energy. A significant contribution of this work was proposed matrix for selection of routing protocols according to the mobility model and performance parameters.

Hybrid routing protocols have been defined more recently. There has therefore been an emphasis on comparing hybrid routing protocols to reactive and proactive ones. However, almost all previous works made use of common performance metrics among which: end-to-end delay [37], [38], throughput [37], [38], [39], PDR [37], [39], and sometimes Normalized Routing Load [36]. Energy related metrics have been considered in very few works like [9]. Besides the well-known performance metrics such as the throughput and the delay, the remaining energy at a node has also been considered by authors. But this energy related metric is not good enough to appreciate the impact of routing protocol on the energy consumption of the network. This because all the nodes do not necessarily have the same energy consumption scheme.

More recent works can be found in [40] and [41]. The work in [40] is tackling also the energy evaluation issue in wireless mesh network. But this work considers only the reactive mode of HWMP. Authors in [41] have evaluated the performance of ad hoc networks under deterministic and probabilistic channel conditions using single and multipath routing protocols. However, they considered only proactive and reactive routing protocols. To the best of our knowledge, none of the previous work has compared the performance of HWMP in reactive and proactive mode, OLSR, and AODV using energy related metrics.

## 4. SIMULATION SETUP

Network Simulator (NS) version 3.25 is used to evaluate the selected routing protocols. It is considered as one of the best network simulation tools [43].

### 4.1. ENERGY MODEL

The energy consumed in this work is the energy used by the WIFI card in its different states. The model used in NS3 to calculate such energy is WifiRadioEnergyModel. In this model, the WIFI card is assumed to be supplied with a voltage of 3 volts. The current used by the WIFI card in its different states (Sleep, Idle, receive, and transmit) can be modified by the user. We do not use the default values of the current which are those of the WIFI card used in {Formatting Citation}. We rather supposed each node being equipped with a PRO/Wireless 3945ABG 802.11a/b/g network card. So, we set the value of the current for the different state of the WIFI card based on the specification of this card [44].

HWMP has been defined for mesh networks, and it considers each node as a mesh device. In NS3, a mesh device is a device possessing multiple WIFI interfaces. So, directly evaluating the energy consumption of a mesh device is not possible in NS3. It has been therefore imperative to defined new functions for this purpose.

### 4.2. PARTICULARIZATION OF ROUTING PROTOCOLS

Since default values may bias the comparison results, several tests have been carried out to set the attributes of routing protocols with values that guaranteed a fair comparison. The concerned attributes for the different protocols are presented in Table 1. Attributes that do not appear here are left with their default values as presented in Doxygen documentation [45].

Table 1. Routing protocols particularization.

| Routing protocol | Parameters | Values |
|---|---|---|
| AODV | HelloInterval | 3s |
| | RreqRetries | 5 |
| | ActiveRouteTimeout | 100 |
| | DestinationOnly | True |
| OLSR | HelloInterval | 3s |
| HWMP | RandomStart | 0.1s |
| | UnicastPreqThreshold | 10 |
| | UnicastDataThreshold | 5 |
| | DoFlag | True |
| | RfFlag | False |

### 4.3. MOBILITY AND PATH LOSS MODELS

Three models are used in this work. The Constant Position Mobility Model is the model used to keep the nodes at constant position during all the simulation. It is the model that was used for all the scenarios that involve static nodes. The Random way mobility Model is the mobility model that defined the mobility of nodes in mobile scenarios. In this model, a node starts moving towards a random waypoint at a random speed then when it reaches the destination it stops choose a new waypoint and a new speed then starts its movement in that new direction. This process is repeated for each node till the end of the simulation.

To calculate how the signal is attenuated, we use Log-distance Propagation Loss Model. This model calculates the reception power with a so-called log-distance propagation model defined by (1).

$$PL = PL_0 + 10n \, log_{10} \left( \frac{d}{d_0} \right) \quad (1)$$

$n$: the path loss distance exponent

$d_0$: reference distance (m)

$PL_0$: path loss at the reference distance (dB)

$d$: distance (m)

$PL$: path loss (dB)

When the path loss is requested at a distance smaller than the reference distance, the Tx power is returned. The default reference loss of 46.6777dB corresponds to reference distance of one meter [45].

### 4.4. TOPOLOGIES AND NODE CONNECTION

We used a grid topology for static position scenarios. The distance between the nodes is set to 180 meters to make sure that with an 802.11g WIFI network card, a node can only forward its packets through the next closest nodes.

For mobile scenarios, the topology boundaries are set depending on the number of nodes. For $N$ nodes, the area on which the nodes move is a square with a length side of $N * 180$ meters. When using HWMP in its proactive mode (HWMP-P), the root node is the node situated at the center of the grid in the case of the static scenarios. For mobile scenarios, the root node is placed at the center of the square area, and remains static. Figure 1 illustrates a 25 nodes placement in both scenarios.

In all the scenarios, the number of connections is equal to three times the grid width and the duration of each connection is randomly generated by an exponential variable. In fact, we have 15 connections for 25 nodes, 21 connections for 49 nodes and 27 connections for 81 nodes. For all the scenarios, the grid width represents the number of CBR stream initiated towards the central node (root node) that we assume having an Internet connection and might serving as a gateway like in the case of an ordinary network. Another number of CBR streams equal to the grid width are initiated with the central node (root node) as the source. Finally, another grid width number of CBR streams are initiated with sources and destination taken randomly among the rest of nodes which are not yet involved in any connection. For 25 nodes for instance, we have 5 connections established between 5 sources randomly chosen, and the central node. We also have 5 connections with the central node as source and 5 nodes taken random as endpoints. Finally, we have five other connections established between 10 nodes (sources/sinks) taken randomly in the rest of the nodes. This connection set-up illustrates a mesh network in which some nodes are communicating through the Internet while others are communicating within the network.

Table 2 contains a summary of the parameters used to carry out all the scenarios.

Table 2.  Simulation parameters.

| Network Simulator parameters | NS3 values |
| --- | --- |
| Topology | Grid topology |
| Mobility Model | ConstantPositionMobilityModel/ Randomwaymobility model |
| Distance between nodes | 180 m |
| Number of nodes | 25, 49 and 81 |
| PHY | MAC 802.11g |
| Propagation loss model | Log-distance Propagation Loss Model |
| Propagation delay model | Constant Speed Model |
| Routing protocol | AODV/HWMP-R/HWMP-P/OLSR |
| Transport protocol | UDP |
| Packet size | 1024 [bytes] |
| Transmission rate | 200[Kbps] |
| Number of connections | 4*Grid Width |
| Connection arrival distribution | Random |
| Data mode | ErpOfdmRate6Mbps |
| Duration of each connection | Exponential (mean = 30s) |
| Sleep current | 0,01[A] |
| Idle current | 0,05 [A] |
| Transmission current | 0,6 [A] |
| Receiving current | 0,467 [A] |
| Simulation time | 180 [sec] |

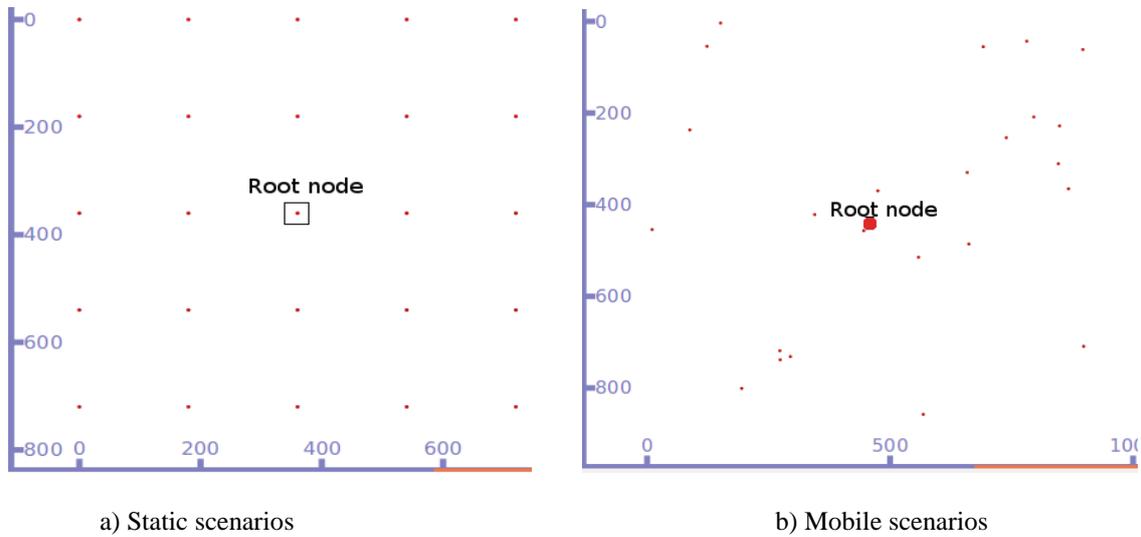

a) Static scenarios          b) Mobile scenarios

Figure 1. Initial positions in scenarios with 25 nodes

## 5. RESULTS AND DISCUSSIONS

### 5.1. ENERGY AND PERFORMANCE

#### 5.1.1. ENERGY CONSUMPTION

Figure 2 enables us to note that, in both static and mobile scenarios, HWMP-P consumed the highest amount of energy. AODV has the lowest energy consumption in both scenarios except in the case of the mobile scenario involving 81 nodes where it consumes a little bit more than OLSR. We also observe that, reactive routing protocols (AODV and HWMP-R) are less influenced by the mobility than proactive routing protocols (HWMP-P and OLSR). We conclude by pointing out that, the layer two routing protocol as defined in the simulator, consumed more energy than layer three routing protocols irrespective of the mobility of the nodes.

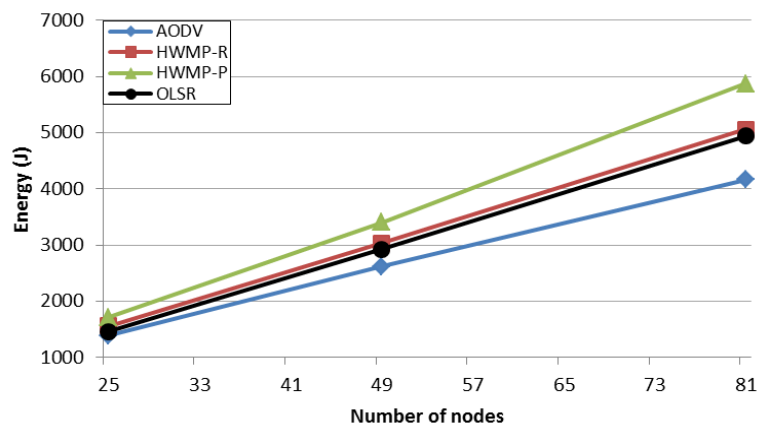

a) Static scenarios

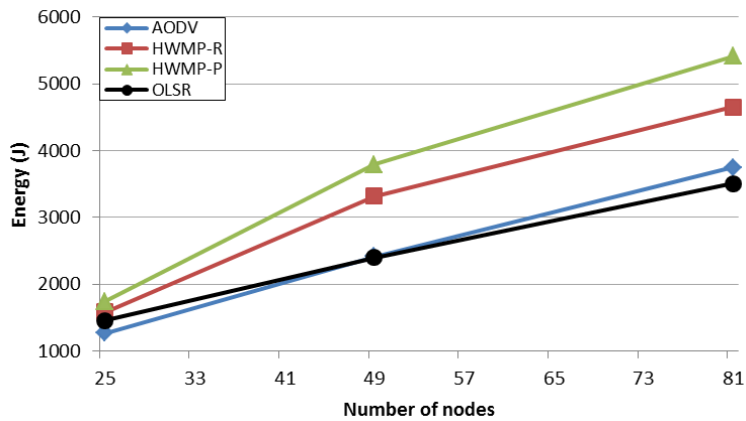

b) Mobile scenarios

Figure 2. Energy consumption measurement

## 5.1.2. THROUGHPUT

Figure 3 shows that AODV is globally the steadiest routing protocol in terms of throughput. In static scenarios, AODV offers the best throughput in the 25 nodes network. For the networks of 49 and 81 nodes, OLSR has the best performance. While the performance of the proactive protocols worsens with the increase in the number of nodes, the performance of reactive protocols grows. We notice that HWMP-R has the worst performance for a number of nodes below 64. We conclude that AODV and HWMP-R are more adapted to network with a high number of nodes while OLSR and HWMP-P are more adapted to network with a few number of nodes.

For the mobile scenarios, proactive routing protocols are highly affected by the mobility of nodes. However, we notice that the effect of mobility affects OLSR more than HWMP-P. The performance of AODV is very steady and grows with the number of nodes. It is important to point out that OLSR has the best performance in static scenarios with the 49 and 81 grid networks and offers the worst performance with the same number of nodes in the mobile scenarios. This shows that the performance of some protocols depends a lot on the topology while other protocols are somehow steady in different scenarios.

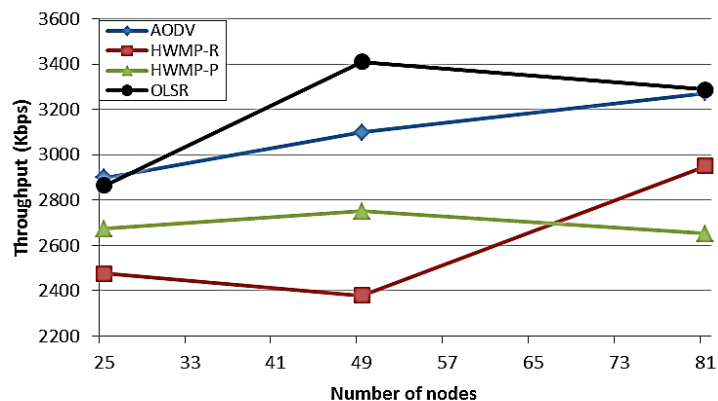

a) Static scenarios

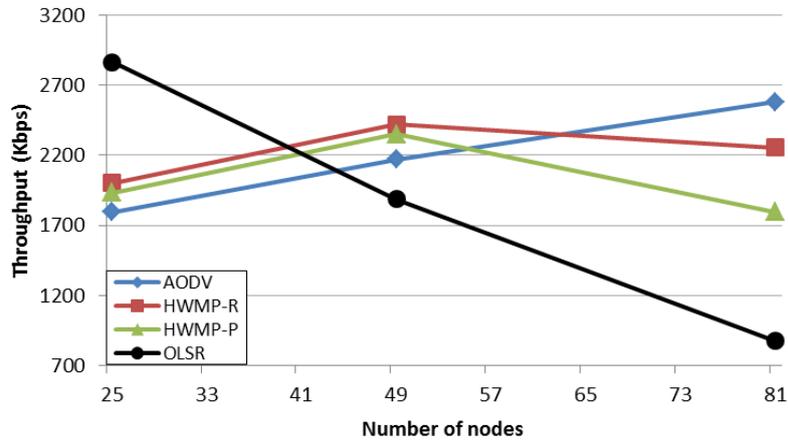

b) Mobile scenarios

Figure 3. Throughput measurement

## 5.1.3. PDR

The values of the PDR in both static and mobile scenarios is given in Figure 4. From this figure, OLSR offers the best performance. We observe that between 25 and 49 nodes AODV has the second-best performance in the static scenario but the worst performance in the mobile scenarios. That shows a very bad mobility impact on the PDR as far as AODV is concerned. In the mobile scenarios in the case of 81 nodes, all the protocols have almost the same performance except HWMP-P which has a PDR around 20%.

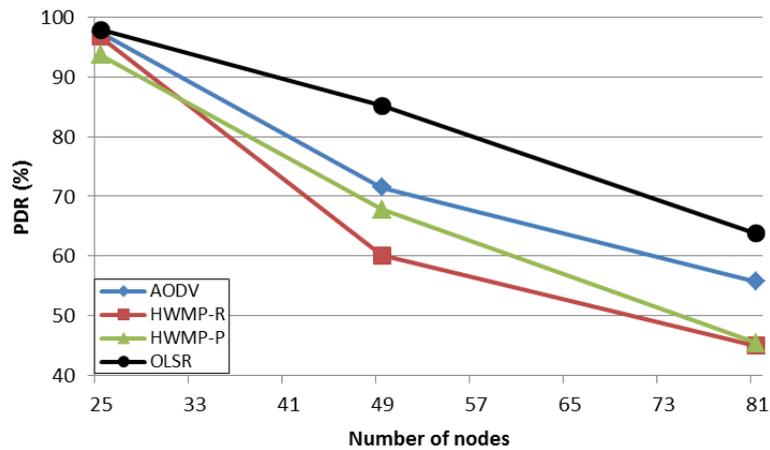

a) Static scenarios

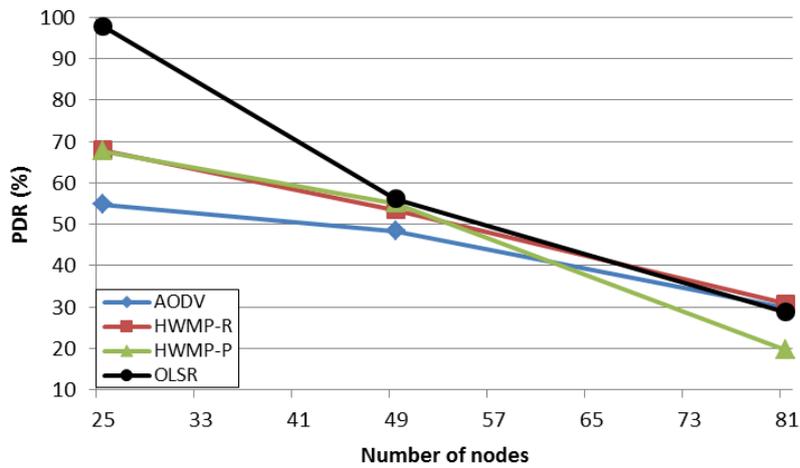

b) Mobile scenarios

Figure 4. PDR measurement

## 5.1.4 Delay

It is obvious from Figure 5 that OLSR has the best delay in all the scenarios. We notice that AODV which has the second-best performance in terms of delay in the static scenarios has the worst delay in the mobile scenarios for a number of nodes above 36. We also notice that the delay of HWMP-R is very close to that of OLSR in the mobile scenarios. Despite the fact that different MAC used, the delay provided by HWMP-R in mobile scenario confirms the result previously found in[9]. For both cases HWMP-P has a delay among the two worst. This makes us to understand that OLSR indifferent scenarios can outperform the other routing protocols regarding delay.

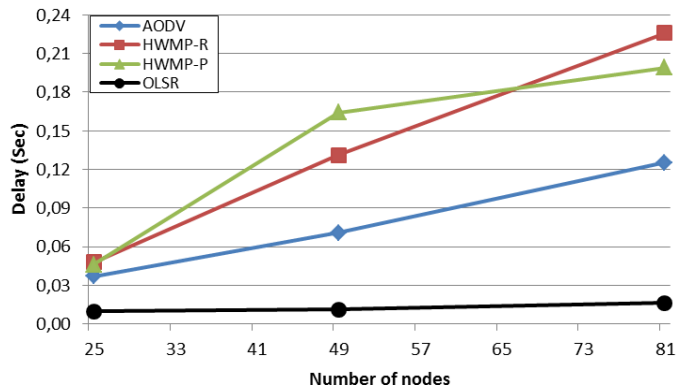

a) Static scenarios

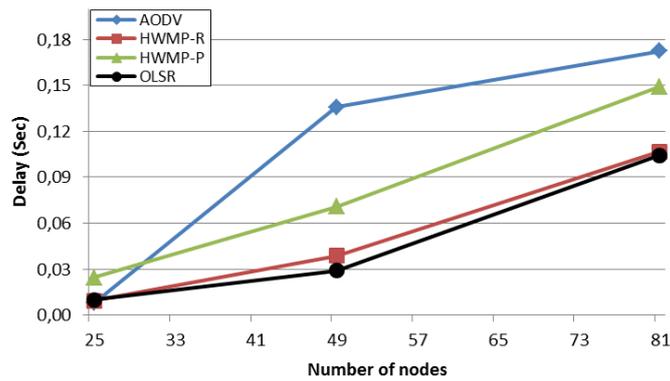

b) Mobile scenarios

Fig. 5. Delay measurement

## 5.2. ENERGY EFFICIENCY

The energy consumption of routing protocols is considered along with other performance metrics in order to avoid biasing the conclusion. The intention is to assess how efficient is the use of energy. Two metrics are therefore used: e-PDR and e-Throughput.

### 5.2.1 E-THROUGHPUT

e-Throughput is defined as the ratio between the energy consumed and the throughput [40]. It is given by equation (2).

$$e - Throughput = \frac{Energy\,consumed}{Throughput} \quad (2)$$

The best protocol is the one with the lowest e-Throughput. Therefore, looking at the bar chart of the static scenarios provided in Figure 6, AODV and OLSR offer almost the same performance for 25 and 49 nodes in terms of e-Throughput. We note that, AODV outperforms all the other protocols in the case of 81 nodes. Regarding the bar chart of the mobile scenarios, except in the case of 25 nodes, where OLSR and AODV have nearly the same performance, the rest of the scenarios are dominated by AODV. We observe that, in almost all the scenarios layer two routing protocols (HWMP-R and HWMP-P) have the worst performance.

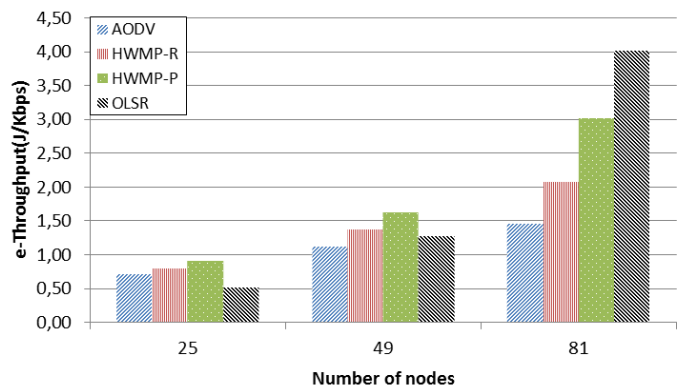

a) Static scenarios

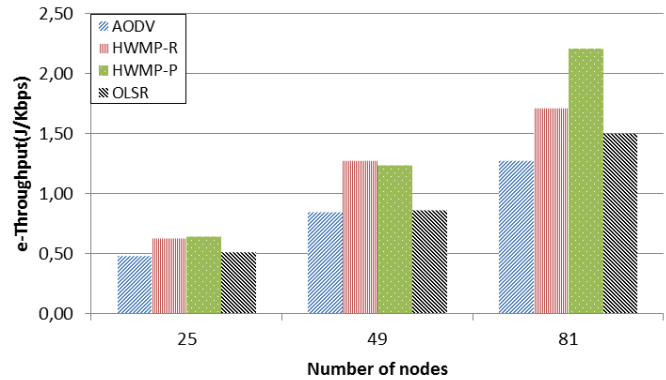

b) Mobile scenarios

Figure 6. e-Throughput measurement

## 5.2.1 E-PDR

e-PDR is defined as the ratio between the energy consumed and the throughput [40]. It is given in equation (3).

$$e - PDR = \frac{Energy\,consumed}{PDR} \quad (3)$$

The routing protocol with the lowest e-PDR, is the protocol that offers the best performance. Figure 7 reveals that, for the static scenarios, AODV and OLSR offer the best e-PDR. We also observe a very slight difference in their performances. HWMP-P and HWMP-R have poor performances in all the scenarios. However, HWMP-P outperforms HWMP-R in almost all scenarios. Therefore, irrespectively of its routing mode, HWMP consumed much more energy than layer three routing protocols.

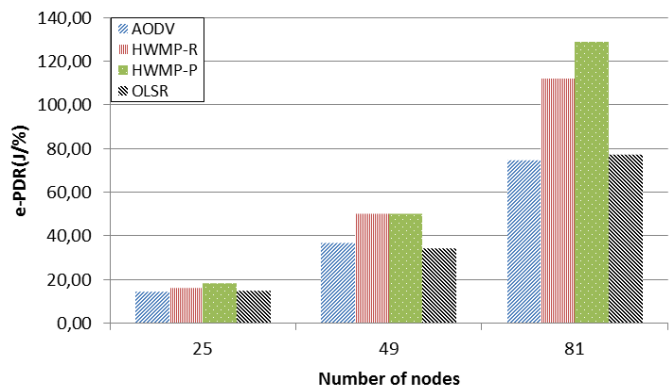

a) Static scenarios

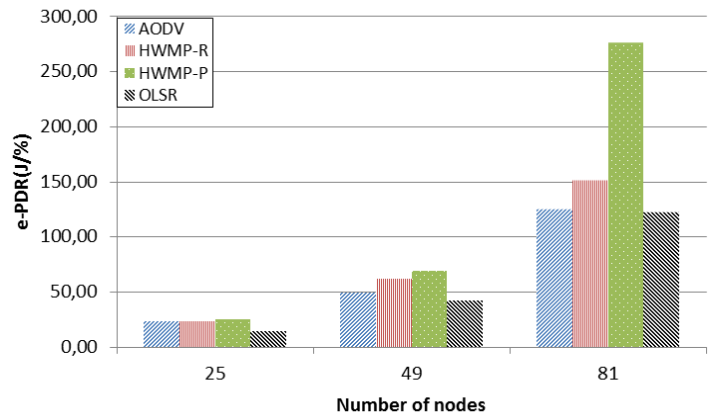

b) Mobile scenarios

Figure 7. e-PDR measurement

# 6. CONCLUSION

In this paper, we aimed at comparing AODV, OLSR and HWMP in its proactive and reactive modes. We used NS3 simulator and the following metrics: energy consumed, throughput, PDR, delay, e-Throughput and e-PDR. We evaluated the routing protocols using two topologies: a grid topology and a mobile nodes topology. The connections between the nodes were established to reflect an Internet access infrastructure. The conclusion we draw in comparing selected routing protocols can be summarized as follow:

- OLSR is globally the most performant routing protocol especially in terms of PDR and delay. However, its throughput can be highly affected by mobility and scalability.
- AODV can offer the same performance as OLSR in several scenarios and seems to be more stable in different network environment than OLSR.
- Regarding not energy-related metrics in dynamic topologies, the performance of HWMP-P and HWMP-R are found between OLSR and AODV. Therefore, HWMP under certain conditions can be useful as a middle solution especially in mobile scenarios.
- Generally, HWMP consumed more energy than AODV and OLSR with usually the worst e-Throughput and e-PDR. However, HWMP-P consumes less energy than HWMP-R.

In our future work, we will look at the impact of multiple root nodes on HWMP, since they can represent a wireless mesh network with multiple gateways. Eventually we shall investigate on an algorithm that is able to give the most adequate position of the root(s) node(s) for a given topology and a number of roots nodes. All these elements will be adjusted together in order to find the best configuration for HWMP.